# THERMAL RADIATION FROM MAGNETIZED NEUTRON STARS
## A look at the Surface of a Neutron Star

DANY PAGE

*Instituto de Astronomía, UNAM, México D.F.*

**Abstract.**
Surface thermal emission has been detected by *ROSAT* from four nearby young neutron stars. Assuming black body emission, the significant pulsations of the observed light curves can be interpreted as due to large surface temperature differences produced by the effect of the crustal magnetic field on the flow of heat from the hot interior toward the cooler surface. However, the energy dependence of the modulation observed in Geminga is incompatible with blackbody emission: this effect will give us a strong constraint on models of the neutron star surface.

## 1. Introduction

Neutron stars (NS) differ from other 'normal' stars in a very peculiar way, besides their intrinsically distinctive physical characteristics: almost all observational data we have about them come from radiation emitted *not* by the star's surface but by its environment. The situation has changed recently due to deep *ROSAT* observations of four nearby NSs for which there is now strong evidence that surface thermal radiation *has* been detected (Ögelman, 1993b). These results finally give us a direct look at the surface of a NS, thus opening a new window on the study of these objects. Despite the low energy resolution of the *ROSAT* PSPC (Position Sensitive Proportional Counter), a large amount of information is contained in these observations and deciphering it can lead to severe constraints on models of the NS surface. These four NSs, PSR 0833-45 (Vela; Ögelman, Finley & Zimmermann, 1993), PSR 0656+14 (Finley, Ögelman & Kiziloğlu, 1992), PSR 0630+178 (Geminga; Halpern & Holt, 1992), and PSR 1055-52 (Ögelman & Finley, 1993), show several common features: a strong soft spectral component at energies below about 0.5 keV, which is interpreted as surface thermal emission and is pulsed at the 10%-30% level, and a much weaker hard tail (except maybe in 0656+14) which in the cases of Geminga and 1055-52 is very strongly pulsed and more than 100° out of phase with the soft component. The pulsed fraction of the soft component is apparently constant in the range 0.08–0.50 keV for Vela and 0656+14, slightly decreasing in 1055-52 and strongly decreasing in Geminga, while the light curve shape is energy independent (below 0.5 keV).

The theoretical study of neutron star surfaces is an extremely challenging physics problem and many important question are still unanswered. One thing however is known for sure about the NS surfaces: they are permeated by enormous magnetic fields (of the order of $10^8 - 10^{14}$ G) which determine almost completely their physical state. Whether the magnetic field induces an abrupt density drop at the surface ('magnetic surface') or/and allows the existence of an atmosphere is still an open question. However, this field also penetrates the crust of the star (if not the whole interior), in particular the upper region (*envelope*) where the heat flow from the hot interior is controlled: if the surface temperature is determined by this heat flow, the anisotropy due to the magnetic field will induce large temperature differences between regions where the field is radial or tangential. These temperature differences are a natural explanation for the observed strong pulsations.

We present here results of a simple modeling of this effect which is able to explain most of the observed characteristics. More important than what we can explain is what we *cannot explain*: we will be looking for the 'smoking gun' of the model, hoping that it will tell us in which direction to move for further research. The technical details of our model are described elsewhere (Page, 1995) and we will focus on its general characteristics, strengths and weaknesses, and summarize the results. We will however begin by presenting in § 2 some simple but important facts which are quite obvious to those who have worked on this problem but, for this very reason, are rarely discussed and thus may be unclear to others.

## 2. The Neutron Star Candidates

The total number of NSs in the Galaxy may be of the order of $10^9$, but only of the order of $10^5$ of them are expected to be active pulsars (Narayan & Ostriker, 1990), of which less than 600 have presently been detected (Taylor, Manchester & Lyne, 1993). When looking for candidates for detection of thermal radiation the number shrinks once more by two orders of magnitude: they must be warm enough, nearby and in a region of low interstellar absorption. Models of NS thermal evolution (Page & Applegate, 1992; Shibazaki & Lamb, 1989) show that in probably less than a million years the surface temperature has dropped so much that the thermal emission is practically undetectable. Out of the 558 known pulsars (as of Spring 1994) 101 have a spin-down age less than $10^6$ yrs, of which 21 are below $10^5$ yrs. However most of these young pulsars are at large distances and if we restrict ourselves to candidates within 2.2 kpc the numbers shrink to 13 and 4 respectively. Interstellar absorption suppresses the received flux at energy $E$ by a factor

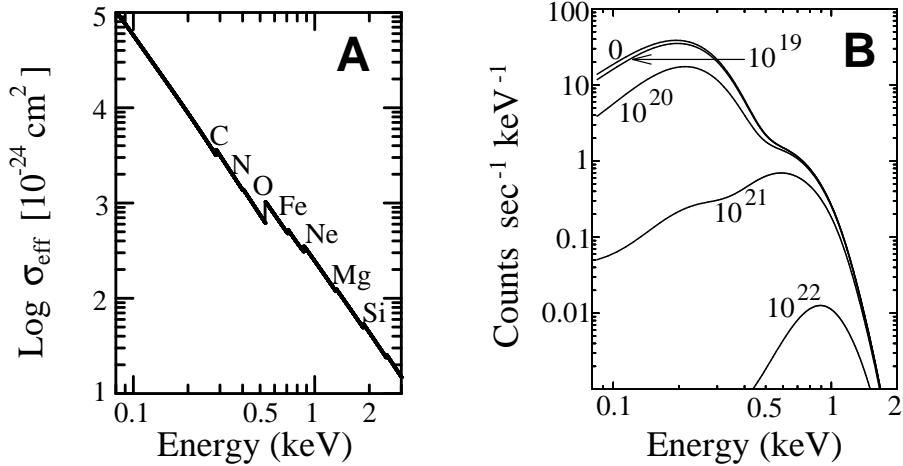

Fig. 1. **A**: Effective cross section for interstellar absorption, taking into account a standard chemical composition of the interstellar medium. Absorption edges from metals are indicated. (From Morrison & McCammon, 1983).
**B**: Blackbody spectrum ($R_\infty = 13$ km, $M = 1.4\ M_\odot$ neutron star at 500 pc with $T_e^\infty = 10^6$ K) and interstellar absorption with various values of $N_H$ as indicated. These spectra take into account the *ROSAT* PSPC response.

$\exp(-N_H \cdot \sigma(E))$, where $N_H$ is the effective neutral hydrogen column density and $\sigma(E)$ the effective cross section at energy $E$ ($\sigma(E) \sim E^{-3}$, Figure 1A). The resulting effect on the received spectrum is illustrated in Figure 1B: values of $N_H$ below $10^{20}$ cm$^{-2}$ have little effect but values around $10^{21}$ cm$^{-2}$ are a serious handicap and a $N_H$ well above $10^{21}$ cm$^{-2}$ leaves little hope for detection of thermal radiation unless the star has a very high temperature ($\gg 10^6$ K) and thus still emits strongly and at high energies where absorption is relatively weak. We show in Figure 2 all known pulsars within 2.2 kpc. It appears that only four NSs are in good observing conditions for our purpose, i.e., young, nearby and within a region of low interstellar density, and these are precisely the four for which surface thermal emission has been certainly detected. Many other young distant or old nearby NSs have been also observed and several detected: only the above four show good evidence for surface thermal emission (Ögelman, 1993b), and in most of the other cases the emission is probably of magnetospheric origin (Ögelman, 1993a). Figure 2 shows clearly that these four NSs will always remain the best objects to study since, even with the use of the next generation of more sensitive detectors, the other candidates will require extremely long observations to collect sufficiently many photons to allow any detailed study. This is a sad fact, but we will have to live (and work) with it.

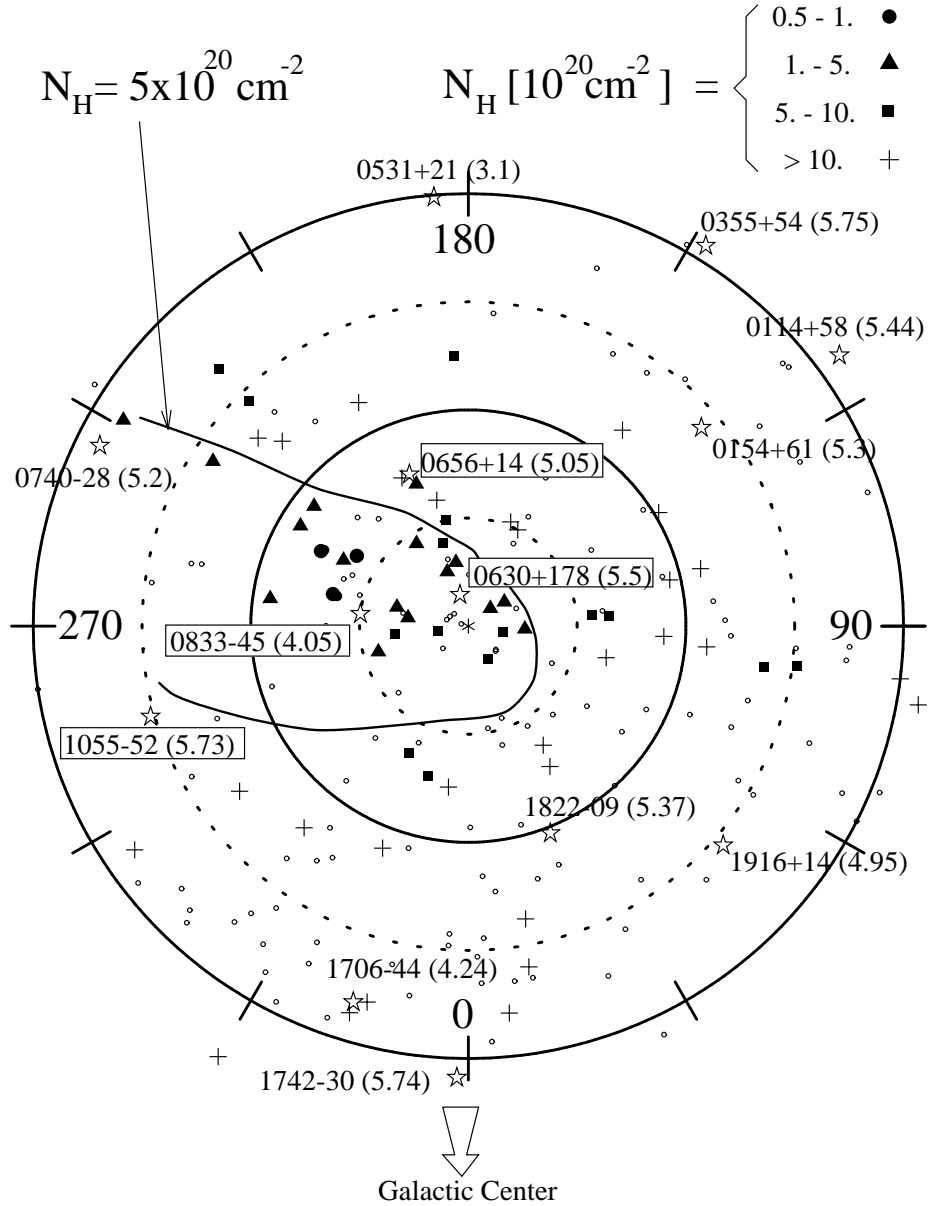

Fig. 2. Projection on the Galactic plane of the position of all known pulsars within 2.2 kpc (small circles). Sun located at center of the figure. Pulsars younger than $10^6$ yrs are marked as stars and individually labeled with the log of their spin-down age given in parenthesis. The four pulsars labeled in a box are those which show surface thermal emission. Other symbols are reference objects to which column densities $N_H$ have been estimated (Frisch & York, 1983) and the continuous line indicates the boundary of $N_H \sim 5 \times 10^{20}\,\mathrm{cm}^{-2}$. These $N_H$ values, as well as distances, are only indicative, and the clumpy nature of the interstellar medium has to be taken into account (see, e.g., the two objects with very different $N_H$ next to PSR 0656+14).

## 3. The Model

Since the only thing known for sure about the neutron star surface is the presence of a strong magnetic field, we developed a minimal model which takes into account the effect of the crustal magnetic field on the surface temperature distribution (Page, 1995). This model is based on the following four postulates:

**A** The soft component detected *is* thermal radiation from the whole surface of the neutron star.

**B** The surface temperature is determined by the heat flow from the NS interior through the magnetized crust (and magnetic field effects on the thermal conductivity can be modeled).

**C** The observed modulation of the light curves is due to the presence of large surface temperature differences (induced by the magnetic field: B).

**D** The surface emits as a blackbody.

The alternatives to these postulates are:

**A'** It may be magnetospheric emission, but no model has been proposed to date.

**B'** The NS surface may be heated from the magnetosphere. One model presented for Geminga (Halpern & Ruderman, 1993) proposes back scattering toward the whole surface of the hard x-rays emitted by the hot polar caps. The expected temperature is of the order of magnitude of what is observed and non-uniform heating of the surface could occur naturally.

**C'** The soft x-ray flux emitted from the surface has to pass through the magnetosphere where resonant scattering and absorption should occur and reshape the light curve (Halpern & Ruderman, 1993).

**D'** The possibilities here are numerous: magnetized atmosphere of pure hydrogen (Shibanov *et al.*, 1992) or heavier elements (Miller, 1992), magnetic surface with no atmosphere at all (Ruderman 1974; Brinkman, 1980), magnetic surface partially covered by a tenuous atmosphere (Fushiki, Gundmundsson & Pethick, 1989), magnetic molecules and chains (Lai, Salpeter & Shapiro, 1992), etc ... Moreover, a patchwork of these possibilities may occur, produced by large differences in temperature and/or magnetic field strength and/or chemical composition over the surface.

Distinguishing between these possibilities can only be done by examining each one in detail and comparing with the observations. Our minimal model A+B+C+D is the simplest one and many of its results also apply to the variant A+B'+C+D. The critical objection is formulated in C': were this reshaping to be significant, we would not be able to learn much about the surface itself. The case D' is the most interesting of all: our main purpose will be to find precisely such a failure (the 'smoking gun') of our minimal model that can tell us what kind of spectrum is appropriate, in which case we could learn directly about the physical state of the surface itself.

### 3.1. Description of a minimal model

Calculations of heat transport in magnetized neutron star envelopes have been performed by several authors (Hernquist, 1985; Van Riper, 1988; Schaaf, 1990a,b) and do predict very large temperature differences at the surface: regions where the field is radial can be up to ten times warmer than regions where the field is tangent to the surface. The temperature difference of course increases with the strength of the field and also increases at lower core temperatures since thermal effects wash out the magnetic field effects. We refer the reader to Yakovlev & Kaminker (1994) for a review of magnetized neutron star envelopes. Using these results one can easily calculate the surface temperature distribution within a one dimensional approximation (Greenstein & Hartke, 1983). The surface temperature $T_s(\theta, \phi)$ depends then on the local angle $\Theta_B$ between the field and the surface normal:

$$T_s(\Theta_B) = \chi(\Theta_B) \times T_s(\Theta_B = 0). \tag{1}$$

All these envelope calculations only include photon and electron transport: in the case of orthogonal transport (i.e., $\Theta_B = 90°$) other mechanisms could be important such as lattice or ion transport, extraordinary photon modes, etc. These corrections are not too important at the present stage of development of the model: the range of variation of temperature predicted by the published models is so large and so strongly dependent on the field strength that inclusion of these corrections can be mocked up by a rescaling of the field strength (which we do not know anyway and is a free parameter). With the steep temperature gradients occuring in the orthogonal case, convection may develop (but seems to be suppressed by the magnetic field: Geppert, 1990) and, possibly, meridional circulation (D. G. Yakovlev, private communication) may also smooth the temperature distribution: from a practical point of view this will only affect regions of the star with small areas and low temperatures which thus contribute little to the observed flux. Finally, at low surface temperature (below about $3 \cdot 10^5$ K) the magnetic effects on the upper layers' equation of state become so large that the model is only illustrative (see, e.g., Van Riper, 1988). Future high quality data and observations of cold neutron stars will certainly require improvements, but for the present situation this model of temperature distribution is reasonably adequate. A detailed description of the model is given by Page (1995).

3.2. RESULTS OF THE MINIMAL MODEL

The effects of temperature non-uniformity on the spectrum (with local black body emission) turn out to be very small, particularly when the detector (*ROSAT*'s PSPC) spectral resolution is taken into account, in which case the spectrum is practically indistinguishable from a blackbody spectrum at the star's effective temperature. The situation may be different when realistic spectra, incorporating the strong anisotropies induced by the magnetic field, are considered.

The shape of the observed light curves appear to be incompatible with a dipolar surface field, a result that is not very surprising (a surprise would be an exactly dipolar surface field). Gravitational lensing flattens the modulations so much that the observed pulsed fractions cannot be reproduced with a dipolar field and blackbody emission, unless low mass and large radii are assumed (an unlikely situation which cannot be excluded a priori). However, non-blackbody spectra can produce larger pulsed fractions, but beaming of the emission along the field results in strong narrow peaks, in contradiction with the data (Yu. A. Shibanov, private communication). Geminga and PSR 1055-52 are considered as orthogonal rotators from the analysis of their $\gamma$-ray and radio emissions, respectively: they both show only one peak in the soft x-ray band which imply that the two magnetic poles must be close to each other, i.e., their surface fields are very different from a dipolar field. Observations of several radio pulsars at high frequency do show anomalies in the dispersion measure (Kuz'min, 1992): since this emission is expected to originate closer to the NS surface than the emission at low frequency, these anomalies are interpreted as a distortion of the field by a quadrupolar component (Davies *et al.*, 1984). Similar high frequency observations of PSR 0656+14, 0833-45 and particularly 1055-52 (Geminga has not been detected at radio wavelengths) would be very interesting for comparison with the x-ray observations.

When considering non-dipolar fields the problem of infinite choices immediately arises. Dipole+quadrupole is of course a first candidate and it can produce higher pulsed fractions, but the orientation of the quadrupole with respect to the dipole must be specially chosen: random orientations almost always result in such complicated surface temperature distributions that the resulting light curves show very little modulation. Since all four candidates show strong pulsations, if the surface field is dipole+quadrupole with very little higher order contribution there must be a good physical reason for the required particular orientation to be present in all four cases.

Several models of neutron star structure predict very small radii, which imply enormous gravitational lensing (maximum lensing angle close to 360°

TABLE I

Goodness (reduced $\chi^2$) of various spectral fits, from Finley *et al.* 1992 (0656+14), Meyer *et al.* 1994 (Geminga) and Ögelman & Finley 1993 (1055-52). Due to differences in methodology values should not be compared between objects. The goodness is limited mostly by uncertainties in the detector calibration.

|         | Blackbody | Mag. H | Non mag. H | Non mag. He | Power law |
|---------|-----------|--------|------------|-------------|-----------|
| 0656+14 | 1.43      | -      | -          | 1.20        | -         |
| Geminga | 1.00      | 1.10   | 0.94       | -           | -         |
| 1055-52 | 0.95      | -      | -          | -           | 0.98      |

are possible) and a dramatic suppression of the light curve modulations. The large observed pulsed fractions raise the hope of constraining such models. However it is possible to take surface field configurations which produce high pulsed fractions even in these cases, as for example an off-centered dipole or two off-centered dipoles with different strengths and orientations. Thus, such constraints on the neutron star size require more information on the surface field.

In short, dipolar surface fields are inadequate, but complex configurations allow one to reproduce almost all observed characteristics: blackbody like spectra; shape of the light curves even with strong gravitational lensing; phase of the peak independent of the photon energy, as observed at energies below 0.5 keV. The minimal model is thus quite successful, we need to invoke neither external heating of the surface (B') nor magnetospheric absorption (C') (which does not mean that these effects are not present). A detailed account of these results can be found in Page (1995) and Page & Sarmiento (1995). There is, however, one significant discrepancy which is considered in the next section.

## 4. The Smoking Gun: The 'Geminga Effect'

The weakest assumption of our minimal model is blackbody emission (D), which is certainly wrong (D'). It is hence important to find some observed feature which can not be reproduced by the model and can guide us toward the correct atmosphere model(s). The spectra offer little hope due to the PSPC limited energy resolution and the calibration uncertainties in the low energy range: fits with various types of spectra give equally good results, as illustrated in Table I. The shape of the light curves can easily be reproduced when considering arbitrary field configurations. However there is one distinctive feature in the Geminga light curves which is *absolutely impossible to reproduce with blackbody emission*: the pulsed fraction in the energy band 0.08–0.28 keV is significantly higher than in the 0.28–0.50 keV band, about 30% vs. 20% (we will call this the 'Geminga effect'). We have tried many different field configurations: centered dipole, off-centered dipole, two dipoles, dipole+quadrupole, warm plate(s) of arbitrary size and temperature. In all cases the pulsed fraction below 0.5 keV *increases* with increasing photon energy. This result is certainly still valid if postulate B is replaced by B' and demonstrates the inadequacy of the blackbody spectrum (as long as postulate C is accepted, of course).

Spectra from *non*-magnetic hydrogen atmosphere (Page, Zavlin, Pavlov & Shibanov, unpublished results) also produce increasing pulsed fractions, but magnetic hydrogen atmosphere spectra (Yu. A. Shibanov, private communication) with dipolar field and the corresponding surface temperature distribution do produce a small decrease in the pulsed fraction. However, in a model inspired from plate tectonics (Ruderman, 1991) the Geminga effect appears very naturally (Page *et al.*, 1995), a result which seems to indicate that the neutron star surface is a patchwork of regions with very different physical conditions (see D' in § 3). It is interesting to notice that, if we consider effective temperatures obtained with blackbody fits, Geminga is the coolest of our four NSs, the next one being PSR 1055-52 which shows only a slight 'Geminga effect', while the two warmer pulsars 0656+14 and 0833-45 seem to have energy independent pulsed fractions.

## 5. Conclusions

There is strong evidence that surface thermal emission has been detected in the soft x-ray band (0.08–0.50 keV) from the four PSRs 0833-45 (Vela), 0656+14, 0630+178 (Geminga), and 1055-52 (see Ögelman, 1993b). The basic properties of this emission, i.e., strong pulsed fractions, shape of the light curves, energy independence of the peak phase, can be easily understood within a simple model which assumes a non-uniform surface temperature distribution produced by the anisotropy of heat transport in the outer crust due to the presence of strong magnetic fields (Page, 1995; Page & Sarmiento, 1995). However, when blackbody spectra are used it is found that the light curve modulation (pulsed fraction) always *increases* with increasing photon energy, a result in strong disagreement with the sharp *decrease* observed in the case of Geminga ('Geminga effect'). This contradiction is a clear indication that blackbody spectra are not adequate for describing neutron star thermal emission; it is apparently the only such observational indication. There are many theoretical reasons for such an inadequacy and this 'Geminga effect' will provide a strong constraint on models of the neutron star surface.

## Acknowledgements

I benefited from many discussions with D. G. Yakovlev, Yu. A. Shibanov, A. Pothekin and A. Sarmiento, which strongly influenced the contents of this paper. This work was supported by a UNAM-DGAPA grant IN105794 and a *Cátedra Patrimonial* from Conacyt.